\documentclass[12pt]{article}
\usepackage{bbm}
\usepackage{latexsym}
\usepackage{epsfig}
\newcommand{\lesssim}{\:\mbox{\raisebox{-3pt}{$\stackrel%
{\displaystyle <}{\sim}$}}\:}
\newcommand{\gtrsim}{\:\mbox{\raisebox{-3pt}{$\stackrel%
{\displaystyle >}{\sim}$}}\:}
\begin{document}
\title{\normalsize \hfill UWThPh-2004-13 \\[1cm] \LARGE
Models of maximal atmospheric neutrino mixing \\
and leptogenesis\footnote{Talk presented by W. Grimus at the 5th
    Workshop on \textit{Neutrino Oscillations and their Origin (NOON2004)},
    February 11--15, 2004, Tokyo, Japan}}
\setcounter{footnote}{2}
\author{Walter Grimus\thanks{E-mail: walter.grimus@univie.ac.at} \\
\small Institut f\"ur Theoretische Physik, Universit\"at Wien \\
\small Boltzmanngasse 5, A--1090 Wien, Austria \\*[3.6mm]
Lu\'{\i}s Lavoura\thanks{E-mail: balio@cfif.ist.utl.pt} \\
\small Universidade T\'ecnica de Lisboa \\
\small Centro de F\'\i sica das Interac\c c\~oes Fundamentais \\
\small Instituto Superior T\'ecnico, P--1049-001 Lisboa, Portugal \\*[4.6mm] }

\date{24 May 2004}

\maketitle

\begin{abstract}
We discuss two extensions of the Standard Model based on the seesaw mechanism
and on non-abelian family symmetry groups $O(2)$ and $D_4$,
respectively. Both models have a twofold-degenerate neutrino Dirac mass matrix
$M_D$, a Majorana mass matrix invariant under a $\mu$--$\tau$ interchange
symmetry and the predictions of maximal atmospheric neutrino mixing and
vanishing mixing angle $\theta_{13}$. Leptogenesis can naturally be
incorporated if $10^{-3} \lesssim m_1 \lesssim 10^{-2}$ eV where $m_1$ 
the mass of the lightest neutrino and if the relevant heavy neutrinos are in
the range $10^{11}$ to $10^{12}$ GeV. The $D_4$ model is more constrained and
leptogenesis requires $m_1$ to be in the vicinity of $4 \times 10^{-3}$ eV.
\end{abstract}

\newpage

\section{Introduction}

The exploration of neutrino masses and mixing has made tremendous progress in
recent years---for a review see, e.g., Ref.~\cite{maltoni}. 
It has turned out that lepton mixing is very
different from quark mixing, with two angles of the $3 \times 3$ mixing matrix
$U$ being large. These angles are the 
atmospheric neutrino mixing angle $\theta_{23}$ with  
$\sin^2 2\theta_{23} > 0.9$ (90\% CL) or 
$\theta_{23} = 45^\circ \pm 9^\circ$, and the solar mixing angle
$\theta_{12}$, which is large but significantly smaller (at the $5\, \sigma$
level) than  $45^\circ$. On the other hand, the third angle 
$\theta_{13}$ is small, with $\sin^2 \theta_{13} \lesssim 0.05$ ($3 \sigma$
level). 
Though the data do not show \emph{compelling} evidence for maximal atmospheric
neutrino mixing, they nevertheless warrant the 
search for models where $\theta_{23} = 45^\circ$ is enforced by symmetries and
the features of small $\theta_{13}$ and large but non-maximal solar mixing
come out in a natural way. 
For a general assessment of neutrino physics with respect to experiment and
theory see Ref.~\cite{smirnov}.

In this report we discuss models which are ``simple'' extensions of
the Standard Model (SM) in the sense that the gauge group remains the same,
but some multiplets of the SM gauge group are added and the seesaw mechanism
\cite{seesaw} is incorporated. 
We consider only the lepton sector, though there might be interesting
relations between mixing in the lepton and quark sectors \cite{minakata}.
Enforcement of exact (or near exact) maximal atmospheric neutrino mixing
suggests to make use of a non-abelian family symmetry group $\mathcal{G}$ 
\cite{wetterich,low}. In this report we discuss two models, one based on
$\mathcal{G} = O(2)$, which we call $\mathbbm{Z}_2$ model \cite{GLZ2}
for reasons to become clear in Section \ref{Z2model}, 
and another one based on $D_4$ \cite{GLD4}, which
is discussed in Section \ref{D4model}. For a review on those models see
Ref.~\cite{ustron03}. In Section \ref{leptogenesis}, we focus
on leptogenesis in these models.

Other non-abelian family symmetries which have been considered are $A_4$
\cite{A4} and  $S_3$ \cite{S3}.
For a general review on models for neutrino masses and mixing see
Ref.~\cite{altarelli}. 

We want to stress the difference between the effect of non-abelian and abelian
family symmetries. Considering any entry of a Yukawa coupling matrix, the
latter ones allow this entry to be either arbitrary or zero. Moreover, 
given any distribution of ``texture zeros'' in fermion mass matrices,
there is always an abelian family symmetry and a scalar sector such that this
distribution of ``texture zeros'' is reproduced by a symmetry
\cite{joshipura}. Though exact maximal atmospheric neutrino mixing cannot be
enforced by abelian family symmetries, interesting models with 
$\theta_{13} = 0$ can be produced \cite{low}.

\section{Framework}

A suitable framework for constructing models with maximal atmospheric neutrino
mixing is given by the SM supplemented by the seesaw mechanism and soft
family-lepton number breaking \cite{GLZ2,GL02}. 
We denote these lepton numbers by $L_\alpha$
with $\alpha = e, \mu, \tau$. 
We allow for an arbitrary number $n_H$ of Higgs doublets. Then the Lagrangian
is given by
\begin{equation}
\mathcal{L} = \cdots
- \left[ \sum_j \left(
\bar \ell_R \phi_j^\dagger \Gamma_j + 
\bar \nu_R {\tilde\phi_j}^\dagger \Delta_j \right) D_L
+ \mbox{H.c.} \right] 
+ \left( \frac{1}{2}\, \nu_R^T C^{-1} \! M_R^* \nu_R + \mbox{H.c.} \right).
\label{L}
\end{equation}
The charged-lepton singlets are denoted by $\ell_R$ and the lepton
doublets by $D_L$. We have three right-handed neutrino singlets $\nu_R$ whose 
Majorana mass matrix $M_R$ must be symmetric. 
The seesaw mechanism assumes that the scale $m_R$ of the mass
eigenvalues of $M_R$ is much higher than the electroweak scale. 
With the vacuum expectation values (VEVs) 
$\langle \phi_j^0 \rangle_0 = v_j$, 
the mass matrix of the charged leptons and the so-called
Dirac mass matrix in the neutrino sector are, respectively, given by 
\begin{equation}\label{massterms}
M_\ell = \sum_j v_j^\ast \Gamma_j \,,
\quad 
M_D = \sum_j v_j \Delta_j \,,
\end{equation}
and the mass matrix of the light neutrinos is given by
the seesaw formula \cite{seesaw}
\begin{equation}\label{Mnu}
\mathcal{M}_\nu = -M_D^T M_R^{-1} M_D.
\end{equation}
The main points in our framework are the following: 
\begin{itemize}
\item
$L_\alpha$-conservation $\Rightarrow$
$\Gamma_j$, $\Delta_j$ diagonal $\forall\, j$;
\item
Soft $L_\alpha$-breaking by terms of dimension three in $\mathcal{L}$, i.e.,
by a non-diagonal $M_R$.
\end{itemize}

Several remarks are at order:
\begin{enumerate}
\item
The theory introduced here is \emph{renormalizable} and all flavour-changing
1-loop amplitudes are finite.
\item
Since the Yukawa coupling matrices are diagonal, the mass matrices 
$M_\ell$, $M_D$ are diagonal as well; therefore, $M_R$ is the only source of
neutrino mixing. 
\item
The soft $L_\alpha$-breaking by $\nu_R$ mass terms occurs 
at the \emph{high} scale $m_R$, nevertheless this theory is in perfect
agreement with the data, as shown in \cite{GL02}.
\end{enumerate}
This last point is connected with an interesting 
non-decoupling property \cite{GL02} in the scalar sector for $n_H > 1$ in the
limit $m_R \to \infty$: 
flavour-changing vertices $\ell^\pm \to \ell^\pm + S^0$,
where $S^0$ is any neutral physical scalar of the theory, do not vanish in
that limit, a property which stems from charged-scalar vertex corrections. As
a consequence, the amplitude of, e.g.,
$\mu^- \to e^- e^+ e^-$ approaches a constant for large $m_R$, suppressed by
a product of four Yukawa couplings; though small, it might be 
within the reach of future experiment. On the other hand, 
amplitudes of $\mu^- \to e^- \gamma$, $Z \to e^- \mu^+$, 
$\tau^- \to \mu^- \mu^- e^+$,\footnote{The amplitude of this process is given
  by a box diagram, not by a vertex correction.} etc., drop like $1/m_R^2$
and are completely inaccessible experimentally. The same happens for
\emph{all} flavour-changing amplitudes if $n_H = 1$. Details can be found in
\cite{GL02}.

\section{The $\mathbbm{Z}_2$ model}
\label{Z2model}

The $\mathbbm{Z}_2$ model of \cite{GLZ2} is a model of the type described in
the previous section, with three Higgs doublets. 
The model is defined via the following family symmetries:
\begin{itemize}
\item[$\rhd$]
$U(1)_{L_\alpha}$ ($\alpha = e,\mu,\tau$);
\item[$\rhd$]
$\mathbbm{Z}_2^{(\mathrm{tr})}$: 
$D_{\mu L} \leftrightarrow D_{\tau L}$, 
$\mu_R \leftrightarrow \tau_R$, 
$\nu_{\mu R} \leftrightarrow \nu_{\tau R}$,  
$\phi_3 \to - \phi_3$;
\item[$\rhd$]
$\mathbbm{Z}_2^{(\mathrm{aux})}: \;
\nu_{eR},\: \nu_{\mu R},\: 
\nu_{\tau R},\: \phi_1,\:
e_R\, \: \mbox{change sign.}$
\end{itemize}
The groups $U(1)_{L_\alpha}$ are the groups associated with the lepton
numbers, broken softly by the Majorana mass terms of $\nu_R$, which also break
the total lepton number $L = \sum_\alpha L_\alpha$. The $\mu$--$\tau$ 
``interchange symmetry'' $\mathbbm{Z}_2^{(\mathrm{tr})}$ is broken by the VEV
of $\phi_3$; 
this breaking is necessary to allow for $m_\mu \neq m_\tau$. The
symmetry $\mathbbm{Z}_2^{(\mathrm{aux})}$ is an auxiliary symmetry which
ensures that $\mathbbm{Z}_2^{(\mathrm{tr})}$ is intact in the neutrino sector
at tree level. The model is dubbed $\mathbbm{Z}_2$ model because of the
$\mathbbm{Z}_2^{(\mathrm{tr})}$ symmetry. 
The Yukawa Lagrangian which follows from the above symmetries is given by 
\begin{equation}\label{LZ2}
\begin{array}{rcl}
\mathcal{L}_\mathrm{Y} & = & 
- y_1 \bar D_{eL} \nu_{eR} \tilde\phi_1  
- y_2 \left( \bar D_{\mu L} \nu_{\mu R} + \bar D_{\tau L} \nu_{\tau R} \right)
\tilde\phi_1 
\\ && 
- y_3 \bar D_{eL} e_R \phi_1
- y_4 \left( \bar D_{\mu L} \mu_R + \bar D_{\tau L} \tau_R \right) \phi_2
\\ &&
- y_5 \left( \bar D_{\mu L} \mu_R - \bar D_{\tau L} \tau_R \right) \phi_3
+ \mbox{H.c.}
\end{array}
\end{equation}
For the problem of obtaining ``naturally'' $m_\mu \ll m_\tau$ with this
$\mathcal{L}_\mathrm{Y}$ see Ref.~\cite{GLmutau}.

As discussed before, an abelian group cannot enforce \emph{exact} 
maximal atmospheric neutrino mixing. 
Indeed the symmetries listed in the beginning of this
section generate a symmetry group which contains a 
non-abelian part because
$U(1)_{L_\mu} \times U(1)_{L_\tau}$ and $\mathbbm{Z}_2^{(\mathrm{tr})}$ 
do not commute and together these symmetries generate $O(2)$. The full
symmetry group is given by \cite{su5}
\begin{equation}
U(1)_{L_e} \times U(1)_{(L_\mu + L_\tau)/2} \times 
O(2)_{(L_\mu - L_\tau)/2} \times \mathbbm{Z}_2^{(\mathrm{aux})}.
\end{equation}
In this equation the lepton numbers associated with the groups are indicated by
subscripts. 

Now the mass matrices in the neutrino sector of the $\mathbbm{Z}_2$ model are
readily found to be 
\begin{equation}\label{massmatrices}
M_D = \mbox{diag}\, (a,b,b) 
\quad \mbox{and} \quad 
M_R = \left( \begin{array}{ccc} m & n & n \\ n & p & q \\ n & q & p
\end{array} \right).
\end{equation}
Note the $\mathbbm{Z}_2^{(\mathrm{tr})}$ symmetry in these matrices, which
via Eq.~(\ref{Mnu}) is transferred to the mass matrix
\begin{equation}\label{MnuZ2}
\mathcal{M}_\nu = 
\left( \begin{array}{ccc} x & y & y \\ y & z & w \\
y & w & z \end{array} \right)
\quad 
(x, \, y, \, z, \, w \in \mathbbm{C})
\end{equation}
of the light neutrinos.

The neutrino mixing matrix is found by the diagonalization procedure
\begin{equation}
U^T \mathcal{M}_\nu U = \hat m = \mbox{diag}\, (m_1,m_2,m_3) 
\end{equation}
with positive neutrino masses $m_j$. The key to the determination of $U$ is
the observation that $(0,-1,1)^T$ is an eigenvector of $\mathcal{M}_\nu$ of
Eq.~(\ref{MnuZ2}), whence we also obtain $m_3 = | z - w |$.
Then one can show that $U$ has the form 
\begin{equation}\label{U}
U = 
\mathrm{diag}\, (1, e^{i\alpha}, e^{i\alpha}) 
\left( \begin{array}{ccc}
\scriptstyle \cos \theta & \scriptstyle \sin \theta & \scriptstyle 0 \\[1mm]
-\frac{\sin \theta}{\sqrt{2}} & \frac{\cos \theta}{\sqrt{2}} 
& -\frac{1}{\sqrt{2}} \\[1mm]
-\frac{\sin \theta}{\sqrt{2}} & \frac{\cos \theta}{\sqrt{2}} 
& \frac{1}{\sqrt{2}}
\end{array} \right)
\mathrm{diag}\, (e^{i\beta_1}, e^{i\beta_2}, e^{i\beta_3})\,. 
\end{equation}
Thus from the mass matrix (\ref{MnuZ2}) we have the following results
\cite{GLZ2}:  
$\theta_{13} =  0^\circ$,
$\theta_{23} = 45^\circ$,
$\theta_{12} \equiv \theta$ is arbitrary, however, without finetuning it will
be large. 
As for the phases, there is no CKM phase since $U_{e3} = 0$; the physical
Majorana phases are given, e.g., by 
$\Delta \equiv 2(\beta_1 - \beta_2)$ and $2(\beta_1 - \beta_3)$. The phase 
$\alpha$ is non-physical and can be absorbed into the charged lepton fields.
There is no prediction for the masses from Eq.~(\ref{MnuZ2}). However, by
convention we always assume $m_1 < m_2$ for the masses involved in solar
neutrino oscillations. 

That the mass matrix (\ref{MnuZ2}) does not predict the neutrino masses,
follows easily from parameter counting. Taking into account that two unphysical
phases can be removed from Eq.~(\ref{MnuZ2}), while preserving its form, we
see that the matrix (\ref{MnuZ2}) has six real physical parameters. On the
other hand, we have three neutrino masses, three mixing angles and three
physical phases, thus Eq.~(\ref{MnuZ2}) has three predictions, namely those
for $\theta_{13}$, $\theta_{23}$ and the CKM phase which vanishes by virtue of
$\theta_{13} = 0^\circ$. Consequently, parameter
counting does not allow further relations.

\section{The $D_4$ model}
\label{D4model}

The $D_4$ model of Ref.~\cite{GLD4} contains the same multiplets as the
$\mathbbm{Z}_2$ model plus two heavy real scalars 
$\chi_{1,2}$. The family symmetry is the discrete group $D_4$ and the pairs
$(D_{\mu L}, D_{\tau L})$, $(\mu_R, \tau_R)$, 
$(\nu_{\mu R}, \nu_{\tau R})$, $(\chi_1, \chi_2)$ transform
according to its 2-dimensional irreducible represenation. The auxiliary
symmetry $\mathbbm{Z}_2^{(\mathrm{aux})}$ here is the same as for the
$\mathbbm{Z}_2$ model, however, we have a 
discrete $\mathbbm{Z}_2^{(\alpha)}$ version of the lepton numbers, such that 
$\mathbbm{Z}_2^{(\mu,\tau)}$ together with the $\mu$--$\tau$ interchange
symmetry generate the $D_4$. All symmetries are broken spontaneously in this
model. For details we refer the reader to Ref.~\cite{GLD4}.

The Yukawa Lagrangian of the $D_4$ model is given by  
\begin{equation}\label{LD4}
\mathcal{L}'_\mathrm{Y} = \mathcal{L}_\mathrm{Y} +
\left[ \frac{1}{2}\, y_\chi\, \nu_{eR}^T C^{-1} 
\left( \nu_{\mu R} \chi_1 + \nu_{\tau R} \chi_2 \right)
+ \mbox{H.c.} \right]
\end{equation}
where $\mathcal{L}_\mathrm{Y}$ is is the Lagrangian of Eq.~(\ref{LZ2}) and
$y_\chi$ is a coupling constant. Moreover, the $D_4$
symmetry admits $\nu_R$ mass terms with non-zero 
masses $\left( M_R \right)_{ee}$ and 
$\left( M_R \right)_{\mu\mu} = \left( M_R \right)_{\tau\tau}$. As shown in
\cite{GLD4}, the VEVs of $\chi_1$ and $\chi_2$ of order $m_R$ are equal up to
very small corrections of order $(100\: \mathrm{GeV}/m_R)^2$, 
which we neglect in the following. 
Taking all the facts of this paragraph
together, after spontaneous symmetry breaking we are lead to a mass matrix
$M_R$ of the right-handed neutrino 
singlets as given by Eq.~(\ref{massmatrices}), except that in the $D_4$ model
we have $q=0$.   

Integrating out the heavy degrees of freemdom, 
below the seesaw scale $m_R$ the $D_4$ model is identical with the 
$\mathbbm{Z}_2$ model apart from $q=0$. 
As demonstrated in Ref.~\cite{GLD4}, from $q=0$ it follows that only the
normal spectrum with $m_1 < m_2 < m_3$ is allowed. Furthermore, $q=0$ 
induces two more relations among the physical quantities, namely the two
Majorana phases are determined by the light neutrino masses and the solar
mixing angle. In particular, we find \cite{GLD4,GLleptogenesis}
\begin{equation}\label{Delta}
\cos \Delta = 
\frac{(m_1 m_2/m_3)^2 - c^4 m_1^2 - s^4 m_2^2}{2 c^2 s^2 m_1 m_2}
\quad \mbox{and} \quad
\left| \left\langle m \right\rangle \right| \equiv |x| = m_1 m_2/m_3.
\end{equation}
In this equation, $s \equiv \sin \theta_{12}$, 
$c \equiv \cos \theta_{12}$ and 
$\left| \left\langle m \right\rangle \right|$ is the effective Majorana mass
appearing in neutrinoless $\beta\beta$-decay (for $x$ see Eq.~(\ref{MnuZ2})).
With the experimental values of the neutrino mass-squared
differences and the solar mixing angle, Eq.~(\ref{Delta}) constraines the
allowed range of the lightest neutrino mass $m_1$; $\cos \Delta \geq -1$ is
only fulfilled \cite{GLD4} for
$3 \times 10^{-3} \: \mathrm{eV} \lesssim m_1 \lesssim 
7 \times 10^{-3} \: \mathrm{eV}$ or
$m_1 \gtrsim 1.5 \times 10^{-2}$ eV.

\section{Leptogenesis in the $\mathbbm{Z}_2$ and $D_4$ models}
\label{leptogenesis}

\begin{figure}[t]
\begin{center}
\epsfig{file=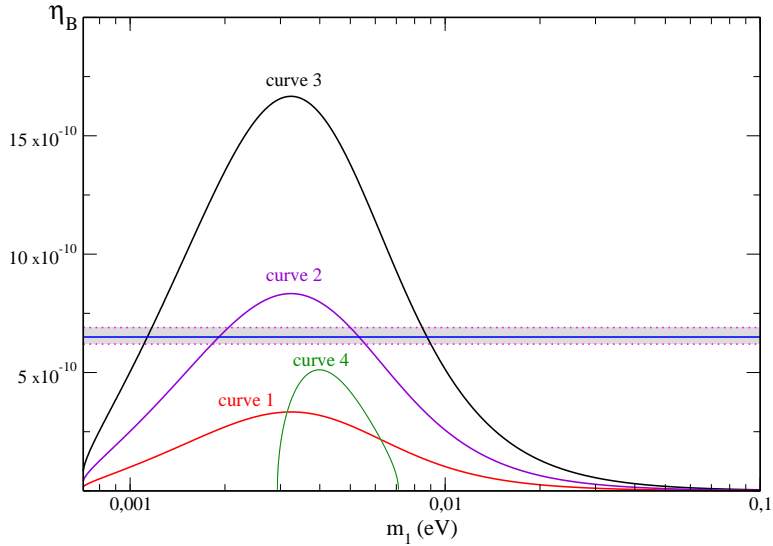,width=9cm,angle=-90} 
\caption{$\eta_B$ as a function of the lightest neutrino mass $m_1$. We have
used as input $\theta_{12} = 33^\circ$, 
$\Delta m^2_\odot = 7.1 \times 10^{-5}$ eV$^2$, 
$M_2/M_1 = 10$, $|v_1| = 50$ GeV and $\Delta = 90^\circ$. 
Curves 1, 2, 3 refer to the $\mathbbm{Z}_2$
model with $M_1 = 1,\, 2.5,\, 5$ in units of $10^{11}$ GeV,
respectively. Curve 4 refers to the $D_4$ model with 
$M_1 = 5 \times 10^{11}$ GeV; its small range in $m_1$ is the effect of
Eq.~(\ref{Delta}). The horizontal lines indicate the experimental value of
$\eta_B$ taken from Ref.~\cite{spergel}. 
\label{etafig}}
\end{center}
\end{figure}
The seesaw mechanism allows to incorporate baryogenesis via leptogenesis
\cite{yanagida}, where the CP asymmetry 
of the decays of the heavy Majorana
neutrinos is responsible that in the present universe we have a tiny fraction
of baryons over photons, denoted by $\eta_B$; for reviews see
Ref.~\cite{lepto-reviews}. 

The suitable basis for the calculation of the CP asymmetry is the physical
basis of the heavy Majorana neutrinos. Defining 
$V^T M_R V = \mbox{diag}\, ( M_1,M_2,M_3 )$, where the $M_j$ are the masses of
the heavy Majorana neutrinos and $V$ is a unitary matrix, 
$V$ has the same 
structure as $U$ of Eq.~(\ref{U}), since $M_R$ and $\mathcal{M}_\nu$ have the
same structure. Thus $V$ is a function of 
$\theta'$, $\chi$, $\gamma_{1,2,3}$ corresponding to 
$\theta$, $\alpha$, $\beta_{1,2,3}$ in $U$, respectively. 
The relevant quantity which appears in the CP asymmetry is 
$\mathrm{Im} \left[ \left( R_{1j} \right)^2 \right]$ 
with the matrix $R$ defined by 
$R \equiv V^T M_D M_D^\dagger V^\ast$ \cite{yanagida,lepto-reviews}. 
In the $\mathbbm{Z}_2$ and $D_4$ models this matrix is given by
\begin{equation}\label{R}
R  = \left( \begin{array}{ccc}
\left| a \right|^2 {c^\prime}^2 + \left| b \right|^2 {s^\prime}^2
&
c^\prime s^\prime \left( \left| b \right|^2 - \left| a \right|^2 \right)
e^{i \left( \gamma_1 - \gamma_2 \right)}
&
0 \\
c^\prime s^\prime \left( \left| b \right|^2 - \left| a \right|^2 \right)
e^{i \left( \gamma_2 - \gamma_1 \right)}
&
\left| a \right|^2 {s^\prime}^2 + \left| b \right|^2 {c^\prime}^2
&
0 \\
0 & 0 & \left| b \right|^2
\end{array} \right). 
\end{equation}
We observe that the third heavy Majorana neutrino does not contribute to
leptogenesis. This is a consequence of the structure of $V$ and of 
$M_D = \mathrm{diag}\, (a,b,b)$ being degenerate.

The $\mathbbm{Z}_2$ model is constrained to such an extent that it allows to
calculate $\theta'$, $2(\gamma_1 - \gamma_2)$, $|a|$, $|b|$ as functions 
of the physical parameters 
\begin{equation}\label{physical}
m_{1,2},\; M_{1,2},\; \theta_{12} \equiv \theta,\; 
\Delta \equiv 2(\beta_1 - \beta_2). 
\end{equation}
With the convention $M_2 > M_1$, 
one can give an explicit analytic expression for the CP asymmetry $\epsilon_1$
of the decay of $N_2$ into $N_1$ in terms of the parameters of
Eq.~(\ref{physical}). Some numerical results for $\eta_B$ are displayed
in Fig.~\ref{etafig}. For the details of the calculation and further numerical
results we refer the reader to Ref.~\cite{GLleptogenesis}.

\section{Summary}

In this report we have revieved the $\mathbbm{Z}_2$ and $D_4$ models of
Refs.~\cite{GLZ2} and \cite{GLD4}, respectively. Both models are simple
extensions of the SM with a non-abelian family symmetry. They predict the
neutrino mixing angles $\theta_{23} = 45^\circ$ and $\theta_{13} = 0^\circ$, 
whereas $\theta_{12}$ remains free but will in general be large. 
The $\mathbbm{Z}_2$ model does
not make any prediction for the neutrino mass spectrum. However, the
requirement of successful leptogenesis forces $m_1$ to be rather small, 
roughly between $10^{-3}$ and $10^{-2}$ eV; this forbids the inverted
spectrum. On the other hand, even without the requirement of successful
leptogenesis, the $D_4$ model predicts the normal
spectrum and leptogenesis restricts the range of $m_1$ further due to
Eq.~(\ref{Delta}). This effect is clearly visible in Fig.~\ref{etafig}.

The twofold degeneracy of $M_D$, imposed by the symmetries of the models,
plays a crucial role for the CP asymmetry in $\eta_B$, which can naturally be
accommodated in the two models. Reproducing $\eta_B$ fixes the orders of
magnitude of both light and heavy neutrino mass spectrum.

\vspace{3mm}

\noindent
\textbf{Acknowledgments:}
W.G. thanks the organizers of NOON2004 for their hospitality and, in
particular, M. Tanimoto for invitation and support.

\end{document}